%% file: draft.tex
\documentclass{IEEEtran}
\usepackage{cite}
\usepackage{amsmath,amssymb,amsfonts}
\usepackage{algorithmic}
\usepackage{graphicx}
\usepackage{textcomp}

\usepackage{algorithmic}

\usepackage[utf8]{inputenc}
\usepackage[T1]{fontenc}
\usepackage{array}
\usepackage{url}
\usepackage{bm}
\usepackage{multirow}
\usepackage{xcolor}
\usepackage{mathtools}
\usepackage{stmaryrd}
\renewcommand{\div}{\operatorname{div}}
\renewcommand{\S}{{\mathbb S}}
\newcommand{\bb}[1]{{\pmb{#1}}}

\newcommand{\R}{{\mathbb R}}

\newcommand{\norm}[1]{{\left\lVert #1 \right\rVert}}

\newcommand{\opd}{\operatorname{d}}
\renewcommand{\div}{\operatorname{div}}
\DeclareMathOperator{\curl}{{curl}}

\renewcommand{\S}{{\mathbb S}}

\DeclareMathOperator{\grad}{grad}

\usepackage{tikz}
\usepackage{pgfplots}
\usepackage{subcaption}

\def\BibTeX{{\rm B\kern-.05em{\sc i\kern-.025em b}\kern-.08em
    T\kern-.1667em\lower.7ex\hbox{E}\kern-.125emX}}
\begin{document}
\title{A Numerical Comparison of an Isogeometric and a Classical Higher-Order Approach to the Electric Field Integral Equation}
\author{ Jürgen~Dölz,~\IEEEmembership{~}
         Stefan~Kurz,~\IEEEmembership{~}
         Sebastian~Schöps,~\IEEEmembership{~}
         Felix~Wolf~\IEEEmembership{~}
\thanks{This work is supported by DFG Grants SCHO1562/3-1 and KU1553/4-1 within the project \emph{Simulation of superconducting cavities with isogeometric boundary elements (IGA-BEM).} Jürgen Dölz is an \emph{Early Postdoc.Mobility} fellow, funded by the Swiss National Science Foundation through the project 174987 \emph{H-Matrix Techniques and Uncertainty Quantification in Electromagnetism}, the Excellence Initiative of the German Federal and State Governments and the Graduate School of Computational Engineering at TU Darmstadt. The work of Felix Wolf is supported by the Excellence Initiative of the German Federal and State Governments and the Graduate School of Computational Engineering at TU Darmstadt. 
All authors are listed alphabetically. Corresponding author: F. Wolf.
}
\thanks{All authors are with the Institute TEMF and the Graduate School of Computational Engineering at TU Darmstadt, 64293, Darmstadt, Germany (e-mails:  doelz@gsc.tu-darmstadt.de, stefan.kurz2@de.bosch.com, schoeps@temf.tu-darmstadt.de, wolf@gsc.tu-darmstadt.de).}}

\maketitle

\begin{abstract}
In this paper, we advocate a novel spline-based isogeometric approach for boundary elements and its efficient implementation. We compare solutions obtained by both an isogeometric approach, and a classical parametric higher-order approach via Raviart-Thomas elements to the solution of the electric field integral equation; i.e., the solution to an electromagnetic scattering problem, promising high convergence orders w.r.t. pointwise error.
We discuss both, the obtained accuracy per DOF, as well as the effort required to solve the corresponding system iteratively, on three numerical examples of varying complexity.
\end{abstract}

\begin{IEEEkeywords}
B-splines, Boundary Element Method, Electric Field Integral Equation, Electric Wave Equation, Isogeometric Analysis, Method of Moments, Raviart-Thomas
\end{IEEEkeywords}

\section{Introduction}

\IEEEPARstart{F}{ollowing} its introduction by Hughes \emph{et al.} \cite{Hughes} isogeometric analysis had a huge leap in popularity.  This holds true especially in electromagnetic applications \cite{Bontinck} and was made possible due to the introduction of isogeometric curl and divergence-conforming discretizations in \cite{IGADIFF}. 
Isogeometric methods enjoy such esteem since they enable users to directly apply geometry representations to avoid meshing errors. By the application of volumetric spline-based discretizations within a Galerkin framework, they offer a better accuracy per degree of freedom (DOF) and smooth solutions, which are often closer to physics, for example, w.r.t. spectral properties, cf. \cite{Cottrell}.

However, commonly only boundary representations are available through CAD applications, and the creation of corresponding volumetric discretizations is non-trivial. 

Thus, recently, as an alternative to isogeometric finite element methods, isogeometric \emph{fast boundary element methods} have been developed, see eg. \cite{Marussig,Dolz} for adaptations for the Laplace and Helmholtz equations. They require only a discretization of the boundary and no additional meshing of the domain. 

Boundary element discretizations in engineering are on the rise since the dense matrices induced by the boundary integral formulation can be handled efficiently by the application of so-called \emph{fast methods}. Most notably among them are the \emph{adaptive cross approximation} \cite{ACA}, the \emph{fast multipole methods} \cite{FMM}, or an approach via \emph{wavelets} \cite{WAVELET}. The latter approach already employed the idea of an exact geometry representation via parametric mappings. All of these approaches have been compared in \cite{COMPARISON} for the lowest order case and scalar problems.

Boundary element methods rely on the existence of \emph{fundamental solutions}, also known as \emph{Green functions}. Herein lies their major restriction, since these functions generally only exist for linear partial differential equations with constant coefficients, with few exceptions.

Within these restrictions lies one application, for which boundary element methods unfold their fullest potential; namely the solution of exterior scattering problems. For these, boundary element methods are exceptionally suited, since they do not require meshing of the unbound exterior domain. 
For acoustic scattering problems, an isogeometric approach to boundary element methods has been discussed in \cite{Dolz}.
For electromagnetic scattering, this specific area of application is well established, often referred to as \emph{method of moments} within the engineering communities. 
They are often applied to solve the exterior \emph{electric wave quation}
\begin{align*}
    \bb\curl \mu^{-1} \bb\curl \bb e - \omega^2 \epsilon \bb e = 0,\qquad\text{ in }\Omega^c
\end{align*}
where $\mu$ denotes the permeability, $ \epsilon$ the dielectric constant and $\omega$ the angular frequency, all assumed to be constant within the domain of interest $\Omega^c$ around the scatterer $\Omega$.
To solve the electric wave equation via a boundary element approach, the unknown is reduced to a vector field on the boundary of the domain, often discretized by divergence-conforming Raviart-Thomas elements, where implementations are presented in \cite{Weggler,BETL}.
An application of this method within the isogeometric framework has first been suggested by Buffa and Vázquez \cite{BuffaProceeding}; and, although first implementations exist, cf. \cite{IGABEM,Simpson}, these have not been compared to classical methods of discretization. This publication aims to close this gap.

In Section \ref{sec::basic} we first introduce the basics of isogeometric analysis, where the NURBS mappings are used for an exact geometry representation. While the isogeometric approach uses B-splines to discretize the unknown by mapping the ansatz functions from the reference domain $\square\coloneqq (0,1)^2$ to (parts of) the physical domain $\Gamma_j$, we will utilize the same method to map Raviart-Thomas elements to the physical domain and compare the different discretizations. By this, we neglect the effect of meshing errors, since the same geometry mappings are used to compare both the spline discretization and the classical Raviart-Thomas discretization. Afterward, we review the problem and introduce the \emph{electric field integral equation.}
After explaining the matrix assembly via a superspace approach in Section \ref{sec::superspace}, we compare both methods of discretization on three different numerical examples in Section \ref{sec::numerics}, after which we draw a conclusion from our findings.

 

\section{A Brief Review of Concepts}\label{sec::basic}

\subsection{Isogeometric Analysis}
    
   Following the framework introduced by \cite{Hughes}, we review the basic notions of isogeometric analysis.
    Let $\mathbb K$ be either $\mathbb R$ or $\mathbb C$. Let $0\leq p< k$. We define a \emph{$p$-open knot vector} as a set
    \begin{align*}
        \Xi = \big[\underbrace{\xi_0 = \cdots =\xi_{p}}_{=0}\leq \cdots \leq \underbrace{\xi_{k}=\cdots =\xi_{k+p}}_{=1}\big]\in[0,1]^{k+p+1},
    \end{align*}
    where $k$ denotes the number of control points.

    We can then define the basis functions $ \lbrace b_i^p \rbrace_{0\leq i< k}$ for $p=0$ as
    \begin{align*}
        b_i^0(x) & =\begin{cases}
            1, & \text{if }\xi_i\leq x<\xi_{i+1}, \\
            0, & \text{otherwise,}
        \end{cases}
        \intertext{ and for $p>0$ via the recursive relationship}
        b_i^p(x) & = \frac{x-\xi_i}{\xi_{i+p}-\xi_i}b_i^{p-1}(x) +\frac{\xi_{i+p+1}-x}{\xi_{i+p+1}-\xi_{i+1}}b_{i+1}^{p-1}(x).
    \end{align*}
    Given the basis as above, the space $S^p(\Xi)$ is given as $\operatorname{span}(\lbrace b_i^p\rbrace_{i\leq k}).$

B-splines on higher dimensional domains are constructed through simple tensor product relationships for controll points $\bb p_{j_1,j_2}  $ via
\begin{align}
      f(x_1,x_2)=\sum_{0\leq i_1< k_1}\sum_{0\leq i_2< k_2}  p_{i_1,i_2} \cdot b_{i_1}^{p_1}(x_1) b_{i_2}^{p_\ell}(x_2),\label{def::tpspline}
\end{align}
which allows \emph{tensor product B-spline spaces}, denoted by 
\begin{align*}
    S_{p_1,p_2}(\Xi_1,\Xi_2),
\end{align*}
to be defined.

    As is often the case in the context of isogeometric analysis, the geometry might be given as a family of smooth, invertible mappings 
    \begin{align}
        \bb F_j\colon \square \to \Gamma_j \subset \mathbb R^3,\label{def::geom}
    \end{align}
       given by NURBS, i.e.,~by
    \begin{align*}
       \sum_{0\leq j_1<k_1}\sum_{0\leq j_2<k_2}\frac{\bb c_{j_1,j_2} b_{j_1}^{p_1}(x) b_{j_2}^{p_2}(y) w_{j_1,j_2}}{ \sum_{i_1=0}^{k_1-1}\sum_{i_2=0}^{k_2-1} b_{i_1}^{p_1}(x) b_{i_2}^{p_2}(y) w_{i_1,i_2}},
    \end{align*}
    for control points $\bb c_{j_1,j_2}$ in $\mathbb R^3$ and weights $w_{i_1,i_2}>0.$
  
    We assume our domains to be boundaries $\partial \Omega$ of some compact Lipschitz domain $\Omega$, and to be parametrized by a family of smooth, invertible NURBS mappings  $\bb F_j\colon  \square\to \Gamma_j.$ 
    We will assume that, for any interface $D = \Omega_j\cap \Omega_i \neq \emptyset$, the mappings coincide, i.e.~that $\bb F_j(\cdot,1) \equiv \bb F_i(\cdot,0)$ holds up to rotation of the reference domain. Moreover, the images of the mappings are assumed not to overlap otherwise.

 
    Let $\bb p$ be a pair of integers $p_1,p_2>0$ and $\Xi_1,\Xi_2$ be $p$-open knot vectors on $[0,1].$ Let $\Xi_j'$ denote their truncation, i.e., the knot vector without its first and last element. We define the spline space $\bb \S^1_{\bb p,\bb\Xi}$ on $ \square$ as 
    \begin{align*}
        \bb \S^1_{\bb p,\bb\Xi}( \square)\coloneqq {}&{}  S^{p_1,p_2-1}(\Xi_1,\Xi_2') \times  S^{p_1-1,p_2}(\Xi_1',\Xi_2).
    \end{align*}
    To define the space in the physical domain, we resort to an application of the pull-backs, which, as a study of \cite{petersonmapped} reveals, is given by
    $
        \iota_1(\bb f_1)\coloneqq \xi \cdot (d\bb F^t)^{-1} (f_1\circ \bb F),
    $
    where the term $\xi$ for $\bb x\in \square$ is given by the so-called \emph{surface measure}
    \begin{align}
        \xi(\bb x)\coloneqq \norm{\partial_{x}\bb F_j( {\bb x})\times \partial_{y}\bb F_j( {\bb x})}_2.
    \end{align}
    In the volumetric cases, the surface measure would coincide with the determinant of the Jacobian.

    Then we define the boundary spline space on $\Gamma_j$ via
    \begin{align}
    \begin{aligned}
        \bb \S^1_{\bb p,\bb\Xi}(\Gamma_j)\coloneqq {}&{} \left\lbrace \bb f  \colon \iota^1(\bb f) \in \bb \S_{\bb p,\bb\Xi}^1( \square)\right\rbrace.\notag
        \end{aligned}
     \end{align} 

\input{data/deRham}

    Proceeding as in \cite{NP}, one can define the corresponding global spline space $\bb \S$. As discussed in \cite{NP}, c.f.~\cite{IGADIFF}, this construction yields globally divergence-conforming discretization, w.r.t. the surface divergence $\div_\Gamma$, if one identifies certain degrees of freedom with each other, to obtain continuity of the normal component across patch interfaces.

    We will use the notation $\bb\S$ and $\bb{\mathcal{RT}}$ if we talk about spline spaces or quadrilateral Raviart-Thomas elements as defined by \cite{Zaglmayer} in general, or when the specific polynomial degrees are clear from context. 
       Otherwise, we will use the notation $\bb\S_p$ and $\bb{\mathcal{RT}}_p$ respectively, to reference the spaces of type $(p,p-1)\times(p-1,p)$.

       Note that this approach is sound, as long as the geometry mappings are smooth. While isogeometric analysis is built such that non-smooth geometry mappings, i.e. NURBS mappings with interior knot repetition, reflect the behavior to discrete space, utilization of such mappings might impact the performance of the $\bb{\mathcal{RT}}$ elements. If non-smooth parts of the mapping overlap with the interior of the Raviart-Thomas elements, approximation properties from the reference domain might not carry over to the physical domain.

       Thus, as test geometries, cf. Fig.~\ref{fig::geoms}, we chose geometries consisting only of rational Bézier patches, i.e. NURBS patches of the same degree in both parameter directions without interior knots. We stress that this is not a limiting factor for isogeometric analysis, cf. \cite{Cottrell}, and not even for a $\bb{\mathcal{RT}}$-based method using geometry mappings, since rational Bézier mapping can easily be extracted from any NURBS mappings, cf.~\cite{Extraction}.

\subsection{Electric Field Integral Equation}
We will now introduce the concepts required for electromagnetic boundary element methods within the scope of this article. For a general introduction to boundary element methods, we refer to \cite{Steinbach}.
To obtain a suitable formulation of the problem, we will first introduce the 
\emph{rotated tangential trace operator}, for smooth functions $\bb u$ given by
\begin{align*}
	\bb\gamma_t(\bb u) = \bb n \times \bb u|_\Gamma,
\end{align*}
and for functions in $\bb H(\bb \curl,\Omega)$ extended via density arguments, cf. \cite{BUH03}. By $\bb n$ we denotes the exterior normal vector of $\Omega$.

We aim to solve the \emph{electric wave equation} under the assumption of constant material coefficients $\mu$ and $\epsilon$ in $\Omega^c$, PEC boundary condition on $\Gamma$ and the Silver-Müller radiation condition \cite{BUH03}.
Fixing an incident wave $\bb g$ we arrive at the equation
\begin{align}
\begin{aligned}
    \bb\curl  \:\bb\curl\: \bb e -\kappa^2 \bb e &= 0,\qquad  \kappa >0\text{ non-resonant,}\label{curlcurl}\\
	\gamma_t(\bb e) &=\gamma_t(\bb g),
\end{aligned}
\end{align}
where, in general, $\kappa \coloneqq \omega\sqrt{\epsilon\mu}.$
Under the assumptions above, it is known that for any solution to \eqref{curlcurl} there exists a surface current $\bb w$ such that the scattered field can be represented by the \emph{electric field integral equation (EFIE)}, c.f.~\cite{BUH03},  given by 
\begin{align}
\begin{aligned}
    \bb e(\bb x) &=-\bb{\tilde{\mathcal V}}(\bb w)(\bb x)\label{efie}
\end{aligned}
\end{align}
with
\begin{align*}
	\bb{\tilde{\mathcal V}}(\bb w)(\bb x)&= \kappa \int_\Gamma G_\kappa(\bb x,\bb y)\bb w(\bb y)\opd \Gamma_{\bb y} \\
    &\qquad+\frac 1\kappa\grad_{\bb x}\int_\Gamma G_\kappa(\bb x,\bb y)\cdot\div_\Gamma\big(\bb w(\bb y)\big)\opd \Gamma_{\bb y},
\end{align*}
for all $\bb x\notin \Gamma.$
The function $G_\kappa$ denotes the Green's function \cite{Dolz}, given by 
\begin{align*}
    G_\kappa(\bb x,\bb y) \coloneqq \frac{e^{i \kappa|\bb x-\bb y|}}{4\pi|\bb x-\bb y|}.
\end{align*}

A variational formulation of \eqref{curlcurl} together with the identity \eqref{efie} makes it possible to obtain the correct surface current required to a representation of the scattered field $\bb e$ via \eqref{efie} by finding a $\bb w \in \bb \gamma_t\big(\bb H(\bb \curl,\Omega)\big)$ such that for
\begin{align*}
	a(\bb w,\bb \phi) \coloneqq \int_\Gamma (\bb\gamma_t\circ\bb{\tilde{\mathcal V}}) (\bb w)
	\cdot(\bb n\times \bb \phi)\opd \Gamma 
\end{align*}
the identity
\begin{align}
	a(\bb w,\bb \phi)=  -
	\int_\Gamma \bb \gamma_t(\bb g)\cdot(\bb n\times \bb \phi)\opd \Gamma\label{varformeq}
\end{align}
holds for all $\bb \phi \in \bb \gamma_t\big(\bb H(\bb \curl,\Omega)\big).$ 
Note that, due to the rotation around the normal, the space $ \bb \gamma_t\big(\bb H(\bb \curl,\Omega)\big)$ need to be discretized in a divergence-conforming way, cf. \cite{NP}.
A discretization of the above yields, that an approximate solution $\bb w_h$ of $\bb w$ is given by the linear system 
\begin{align}
	 \bb A_h\bb w_h = \bb g_h ,\label{eq::naiveSystem}
\end{align} where the matrix entries can be obtained via the formula 
\begin{align*}
&\bb A_{h,i,j}=\\
&{}-\kappa\iint_\square G_\kappa\big(\bb F_i(\bb s),\bb F_j(\bb t)\big){\bb b_j}(\bb s)^{\intercal}d\bb F_i(\bb s)^\intercal d\bb F_j(\bb t){\bb b_i}(\bb t)\opd\bb t\opd\bb s\\
&{}+\frac{1}{\kappa}\iint_\square G_\kappa\big(\bb F_i(\bb s),\bb F_j(\bb t)\big)\div_\Gamma{\bb b_j}(\bb s)\div_\Gamma{\bb b_i}(\bb t)\opd\bb t\opd\bb s,
\end{align*}
see \cite{petersonmapped}. Similarly, one can represent the right hand side via
\begin{align*}
	\bb g_{h,i}=
	&{}-\int_\square (\bb n\times \bb g(\bb F_i(\bb s))) \cdot d\bb F_i(\bb s)^\intercal\bb b_i(\bb s)\opd\bb s. 
\end{align*}
The functions $\bb b_i$, $\bb b_j$ are either isogeometric basis functions from $\bb \S$ or quadrilateral Raviart-Thomas elements $\bb{\mathcal {RT}}$.
Note that, due to the non-locality of $G_\kappa$ the matrix becomes densely populated, thus establishing the need for the already mentioned fast-methods.

\section{The Superspace Approach}\label{sec::superspace}

\begin{figure*}[ht]\centering
	\input{data/nonsmooth} \qquad \input{data/smooth}
	\caption{Superspace-based approach. The right basis can be represented within the left basis by linear combination of ansatz functions. The left basis corresponds to Bernstein polynomials, rescaled in between the knots used to define the spline space on the right hand side.}\label{fig::superspace}
\end{figure*}

We assume the mesh, on which the space $\bb{\mathcal{RT}}$ will be given, be induced by the knot vectors of the isogeometric space. For the construction of a system \eqref{eq::naiveSystem} for either choice of basis functions, we employ a projection based approach, built upon a space $\mathcal P$ of local tensor product polynomial basis $(\bb \pi_j)_{0\leq j<p}$ of the maximum order $p$, defined on every mesh element. Since B-Splines and Raviart-Thomas elements are locally polynomial, we can represent any basis function from $\bb \S$ or $\bb{\mathcal{RT}}$ within $\mathcal P$, as depicted in Fig.~\ref{fig::superspace}. Note that $\bb S\subseteq\bb{\mathcal{RT}}$ holds for all $p$, i.e., every function in $\bb\S$ is representable by a linear combination of functions in $\bb{\mathcal{RT}}$. Note that this approach shares its core ideas with the idea of Bézier extraction for efficient geometry evaluation, cf.~\cite{Extraction}. We will use this to construct the basis functions of either space, similar as in \cite{Dolz}.

Assume $\bb b_j$ and $\bb b_i$ to be basis functions in $\bb\S$ or $\bb{\mathcal{RT}}$. Since the l.h.s. of the problem \eqref{eq::naiveSystem} is induced by the bilinear form $a$, it is clear that, for suitable index sets $I$ and $J$, one finds
\begin{align*}
    a(\bb b_j ,\bb b_i)& = a(\sum_{j'\in J} c_{j'}\bb \pi_{j'},\sum_{i'\in I}c_{i'}\bb \pi_{i'}) \\
    &=  \sum_{j'\in J}c_{j'} a( \bb \pi_{j'},\sum_{i'\in I}c_{i'}\bb \pi_{i'}),\\
    &= \sum_{i'\in I}c_{i'}\sum_{j'\in J}c_{j'} a( \bb\pi_{j'}, \bb \pi_{i'}),
\end{align*}
which, in terms of linear algebra, corresponds to a basis transformation given by application of a sparse transformation matrix $\bb T\in \R^{\ell\times k}$. Hereby $\ell$ denotes the number of DOFs of $\S$ or $\bb{\mathcal{RT}}$, respectively. Similarly, one can transform the r.h.s. of \eqref{varformeq}
thus arriving at a linear system,
\begin{align}
    \bb T \bb A^*_h \bb T^\top \bb w_h = \bb T \bb g^*, \label{eq::notsonaivesystem}
\end{align}
equivalent to the one given in \eqref{eq::naiveSystem}.
This is a known technique in conjunction with fast methods. Since the supports of the $\bb \pi_i$ are highly local, the interaction between clusters of basis functions is diminished, and thus the matrix offers better compression properties, cf. \cite{WegglerMatrix}. However, often it is merely filled with values of $1$ or $-1$, to achieve continuity of the classical Raviart-Thomas elements, while we also introduce smoothness of the spline basis via suitable coefficients.
Using the local tensor product polynomials $\bb \pi_j$ on quadrilaterals in the reference domain $ \square$, we assemble projection matrices $\bb T_{\text{IGA}}$ and $\bb T_{\text{RT}}$ for both the isogeometric basis functions and the Raviart-Thomas elements via local interpolation. At this point, we also introduce the normal continuity across patch interfaces required for the discretization to be divergence conforming. We merely add functions whose DOFs must be identified with each other in the interpolation step. Note that, due to different orientations of patches, a change in sign might be required. 

Thus we can compute the correct index sets $I,J$ together with the correct $c_i$ for our choice of discretization. Note that the unknown vector $\bb w_h$ remains unchanged. A naive pseudo code representation of the assembly of $\bb T$ is as in Fig.~\ref{algo::1}.
\begin{figure}
\begin{algorithmic}
 \STATE {Patchwise $\div$-conforming basis $\lbrace b_i\rbrace $ of size $L$ given}
 \STATE {skiplist$ = [$~$]$;}
 \FOR{$i = 0 \dots L$} 
 \IF{$i\notin\text{skiplist}$} 
 \STATE{$\text{coeffs}=\text{interpolateIn}\mathcal{P}(b_i)$};
 \FOR{$j = 0\dots\operatorname{dim}(\mathcal P)$}
 \STATE{$\bb T(i,j)=\text{coeffs}(j)$;}
 \ENDFOR
 \IF{$b_i$ has normal component at patch interface}
 \STATE{${\text{tmp}} = \text{findIdxOfPartnerFunction}(b_i)$}
 \STATE{$\text{skiplist} = [\text{skiplist},\text{tmp}]$}
 \STATE{$\text{tmpcoeffs}=\text{interpolateIn}\mathcal{P}(b_{\text{tmp}})$}
 \STATE{\text{dir} = \text{findOrientation}($b_i$,$b_{\text{tmp}}$)}\\
 \FOR{$j = 0\dots\operatorname{dim}(\mathcal P)$}
 \STATE{$\bb T(i,j)=\text{tmpcoeffs}(j)\cdot\text{dir}$;}
 \ENDFOR
 \ENDIF
 \ENDIF
 \ENDFOR
\end{algorithmic}
\caption{Algorithm for assembly of $\bb T$. Note that findOrientation returns -1 or 1, depending on the unit direction of the fields in the physical domain. It might be that a vector field needs to be ``glued'' with a negative coefficient to achieve continuity of the normal component across patch interfaces.}\label{algo::1}
\end{figure}

The approach yields dense matrices, as is clear by the representation of the $\bb A_{h,i,j}$ since the Green function does not vanish. For this, we utilize a modified fast multipole method for compression of the matrix $\bb A^*_h$ assembled w.r.t. the functions $\bb\pi_i$, which is explained and analyzed in detail in \cite{IGABEM}. Thus, the error induced by compression of the system matrix equals for both, the Raviart-Thomas as well as the spline-based approach.

\begin{figure*}
\begin{subfigure}[c]{0.3\textwidth}
    \includegraphics[width =  \textwidth]{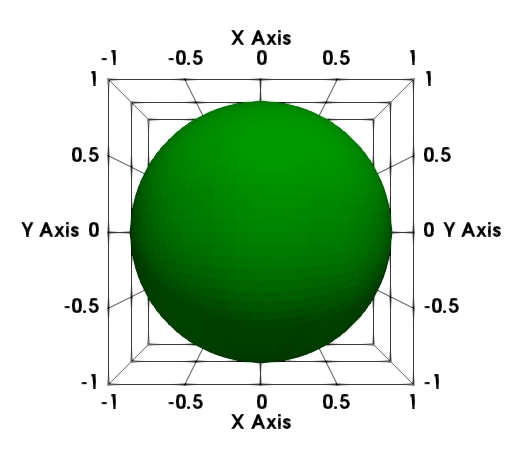}
    \subcaption{Sphere geometry}
\end{subfigure}
\begin{subfigure}[c]{0.3\textwidth}
    \includegraphics[width =  \textwidth]{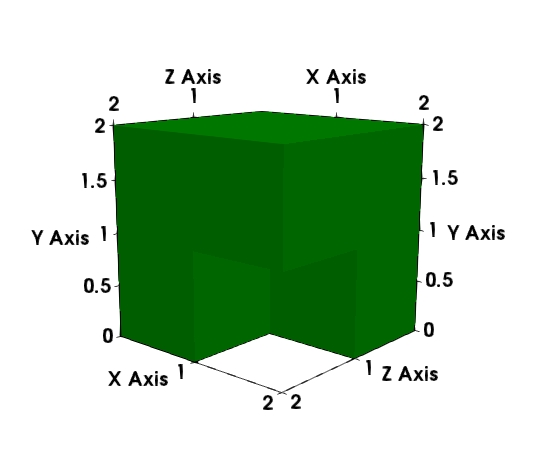} 
    \subcaption{Fichera geometry}
\end{subfigure}
\begin{subfigure}[c]{0.33\textwidth}
    \includegraphics[width =  \textwidth]{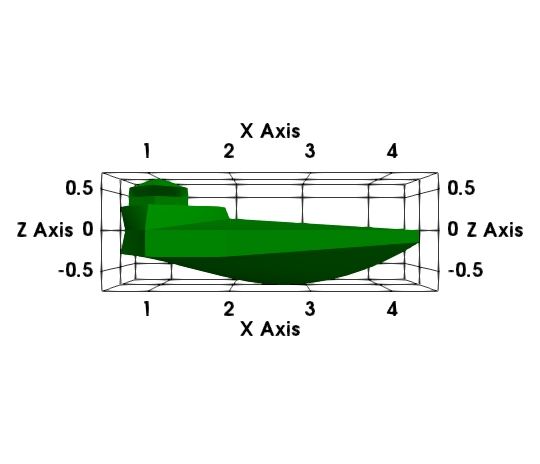}
    \subcaption{Toy Boat geometry}
\end{subfigure}
    \caption{The three different test geometries. For a top view of the Toy Boat geometry see Fig.~\ref{fig::boatpic}}\label{fig::geoms}
\end{figure*}

\section{Numerical Examples}\label{sec::numerics}
We will compare results on three different geometries, depicted in Fig.~\ref{fig::geoms}.
The code used for computation is an improved version of the implementation proposed in \cite{Dolz}. 
We are interested in a comparison of the computational effort required to reach a given accuracy of the quantity of interest $\bb e$. Thus we apply an approach via a \emph{manufactured solution}, whose idea is as follows.

We let $\bb p_0 = [0,0.1,0.1]^\top$ and a non-resonant wavenumber $\kappa >0$ be given and place a Hertz dipole given by  
\begin{align*}
e^{i\kappa r}
\bigg(
\frac{\kappa^2}{r}
(\bb n\times \bb p_0)\times\bb n
+
\bigg(
\frac{1}{r^3}-\frac{i\kappa}{r^2}
\bigg)
\big( 3\bb n(\bb n\cdot \bb p_0)- \bb p_0\big)
\bigg),
\end{align*}
with $r=\|\bb x-\bb x_0\|$ and $\bb n=(\bb x-\bb x_0)/r$
at a point $\bb x_0$, cf. \cite{Jackson}.
One can check that the dipole induces a solution to \eqref{curlcurl} within the domain, either interior or exterior, together with the Silver-Müller radiation condition at $\infty$, not containing $\bb x_0$. By existence and uniqueness of the solution \cite{BUH03}, we know that an exterior evaluation of the solution to \eqref{eq::naiveSystem} via the EFIE must converge to the field induced by the dipole. Moreover, assuming analogy of our approach to other boundary element methods, we can expect increased orders of convergence, cf.~\cite{Dolz}. A proof applicable for the case of the EFIE is provided in \cite{IGABEM}.


Note that, by the construction of the discrete spaces, the lowest order spline space coincides with the lowest order Raviart-Thomas space. This is showcased by the results depicted in Figures \ref{fig::sphere}, \ref{fig::fichera}, and \ref{fig::boat}. Moreover, in addition to the plots of all experiments, detailed numerical data of selected simulations are depicted in Table \ref{examples}.

\subsection{Example I: Unit Sphere Test}
The sphere geometry is given by six NURBS patches as in \cite{Dolz}, where a dipole with wavenumber $\kappa = 1$ is placed at $[0.1,0.1,0]^\top.$
We visualize the maximum pointwise error of 100 evaluations on a sphere with radius 3 around the origin.

Note that, due to the smooth geometry, the effect of higher-order approaches in terms of convergence orders up to $\mathcal O(h^9)$ is clearly visible, see Fig.~\ref{fig::sphere}, and analogous to known results from boundary element theory of acoustic problems \cite[Cor.~3.4]{Dolz}.

For the same level of refinement the Raviart-Thomas discretization yields better accuracies. This is due to the fact that, for the same polynomial degrees and the same level of refinement, the spline discretization is contained in the corresponding Raviart-Thomas space.

However, as can be seen in Figures \ref{fig::sphere}, the spline spaces yield a higher accuracy per degree of freedom, and fewer iterations of the GMRES are required to solve the corresponding system.
\begin{figure*}[!t]
\centering
    \begin{tikzpicture}[scale = .6]
        \begin{axis}[
        height = 7cm,
        ymode=log,
        xlabel=level of refinement,
        xtick={1,2,3,4,5},
        grid = major,
        ylabel=max pw. error,
        legend pos=outer north east
        ]
        \addplot[line width = 1.5pt,blue,mark = triangle*,mark size=3pt] table [trim cells=true,x=M,y=B2] {data/masterBsphere};
        \addplot[line width = 1.5pt,densely dashed,brown,mark = triangle* ,mark options=solid,mark size=1pt]table [trim cells=true,x=M,y=RT2] {data/masterRTsphere};


        \addplot[line width = 1.5pt,red,mark = square*] table [trim cells=true,x=M,y=B3] {data/masterBsphere};
        \addplot[line width = 1.5pt,densely dashed,red,mark = square* ,mark options=solid] table [trim cells=true,x=M,y=RT3] {data/masterRTsphere};
         
        \addplot[line width = 1.5pt,gray,mark = diamond*] table [trim cells=true,x=M,y=B4] {data/masterBsphere};
        \addplot[line width = 1.5pt,densely dashed,gray,mark = diamond* ,mark options=solid] table [trim cells=true,x=M,y=RT4] {data/masterRTsphere};


        \addplot[line width = 1.5pt,purple,mark = *] table [trim cells=true,x=M,y=B5] {data/masterBsphere};
        \addplot[line width = 1.5pt,densely dashed,purple,mark = * ,mark options=solid] table [trim cells=true,x=M,y=RT5] {data/masterRTsphere};
        \end{axis}
        \end{tikzpicture}\qquad
    \begin{tikzpicture}[scale = .6]
        \begin{axis}[
        height = 7cm,
        ymode=log,
        xmode=log,
        xlabel=number of DOFs,
        grid = major,
        ylabel=max pw. error,
        legend pos=outer north east
        ]
        \addplot[line width = 1.5pt,blue,mark = triangle*,mark size=3pt] table [trim cells=true,x=DOF2,y=B2] {data/masterBsphere};
        \addplot[line width = 1.5pt,densely dashed,brown,mark = triangle* ,mark options=solid,mark size=1pt] table [trim cells=true,x=DOF2,y=RT2] {data/masterRTsphere};


        \addplot[line width = 1.5pt,red,mark = square*] table [trim cells=true,x=DOF3,y=B3] {data/masterBsphere};
        \addplot[line width = 1.5pt,densely dashed,red,mark = square* ,mark options=solid] table [trim cells=true,x=DOF3,y=RT3] {data/masterRTsphere};
                \addplot[line width = 1.5pt,gray,mark = diamond*] table [trim cells=true,x=DOF4,y=B4] {data/masterBsphere};
        \addplot[line width = 1.5pt,densely dashed,gray,mark = diamond* ,mark options=solid] table [trim cells=true,x=DOF4,y=RT4] {data/masterRTsphere};


        \addplot[line width = 1.5pt,purple,mark = *] table [trim cells=true,x=DOF5,y=B5] {data/masterBsphere};
        \addplot[line width = 1.5pt,densely dashed,purple,mark = * ,mark options=solid] table [trim cells=true,x=DOF5,y=RT5] {data/masterRTsphere};
        \end{axis}
        \end{tikzpicture} \qquad
        \begin{tikzpicture}[scale = .6]
        \begin{axis}[
        height = 7cm,
        ymode=log,
        xmode=log,
        xlabel=GMRES iterations,
        grid = major,
        ylabel=max pw. error,
        legend pos=outer north east
        ]
        \addplot[line width = 1.5pt,blue,mark = triangle*,mark size=3pt] table [trim cells=true,x=ItB2,y=B2] {data/masterBsphere};
        \addlegendentry{$p=1$ B-spline}
        \addplot[line width = 1.5pt,densely dashed,brown,mark = triangle* ,mark options=solid,mark size=1pt] table [trim cells=true,x=ItRT2,y=RT2] {data/masterRTsphere};
        \addlegendentry{$p=1$ RT}


        \addplot[line width = 1.5pt,red,mark = square*] table [trim cells=true,x=ItB3,y=B3] {data/masterBsphere};
        \addlegendentry{$p = 2$ B-spline}
        \addplot[line width = 1.5pt,densely dashed,red,mark = square* ,mark options=solid] table [trim cells=true,x=ItRT3,y=RT3] {data/masterRTsphere};
        \addlegendentry{$p = 2$ RT} 
        \addplot[line width = 1.5pt,gray,mark = diamond*] table [trim cells=true,x=ItB4,y=B4] {data/masterBsphere};
        \addlegendentry{$p=3$ B-spline}
        \addplot[line width = 1.5pt,densely dashed,gray,mark = diamond* ,mark options=solid] table [trim cells=true,x=ItRT4,y=RT4] {data/masterRTsphere};
        \addlegendentry{$p=3$ RT}


        \addplot[line width = 1.5pt,purple,mark = *] table [trim cells=true,x=ItB5,y=B5] {data/masterBsphere};
        \addlegendentry{$p = 4$ B-spline}
        \addplot[line width = 1.5pt,densely dashed,purple,mark = * ,mark options=solid] table [trim cells=true,x=ItRT5,y=RT5] {data/masterRTsphere};
        \addlegendentry{$p = 4$ RT}
        \end{axis}
        \end{tikzpicture} 
        \caption{Sphere example, $\bb x_0 = [0.1,0.1,0]^\top$, $\kappa = 1$.}    
\label{fig::sphere}
\end{figure*}

\subsection{Example II: Fichera Geometry}
As a geometry, we now employ the Fichera geometry from \cite{Dolz}, given by 24 square patches of length $0.5$, giving the Fichera cube a maximal edge length of 2. The dipole was placed at $\bb x_0 = [0.5,0.5,0.5]^\top$ and, once again, $\kappa = 1$. The evaluation points have been chosen as in the sphere example. Note that the geometry is non-smooth one cannot expect high orders of convergence. 
However, one can still observe an increase in accuracy per DOF, when higher order basis functions are utilized.

Analogously to the previous example, the Raviart-Thomas elements yield higher accuracies w.r.t.~the same level of refinement, whereas the B-splines yield higher accuracies per DOF, see Fig.~\ref{fig::fichera}. Note that the difference in iterations required for a certain accuracy is even greater than in the sphere example. This might be attributed to the fact, those non-smooth geometries with sharp angles yield, in general, badly conditioned systems, compared to those of smooth geometries. 


\begin{figure*}[!t]
\centering
    \begin{tikzpicture}[scale = .6]
        \begin{axis}[
        height = 7cm,
        ymode=log,
        xlabel=level of refinement,
        xtick={1,2,3,4,5},
        grid = major,
        ylabel=max pw. error,
        legend pos=outer north east
        ]
        \addplot[line width = 1.5pt,blue,mark = triangle*,mark size=3pt] table [trim cells=true,x=M,y=B2] {data/masterBfichera};
        \addplot[line width = 1.5pt,densely dashed,brown,mark = triangle* ,mark options=solid,mark size=1pt] table [trim cells=true,x=M,y=RT2] {data/masterRTfichera};


        \addplot[line width = 1.5pt,red,mark = square*] table [trim cells=true,x=M,y=B3] {data/masterBfichera};
        \addplot[line width = 1.5pt,densely dashed,red,mark = square* ,mark options=solid] table [trim cells=true,x=M,y=RT3] {data/masterRTfichera};
        \addplot[line width = 1.5pt,gray,mark = diamond*] table [trim cells=true,x=M,y=B4] {data/masterBfichera};
        \addplot[line width = 1.5pt,densely dashed,gray,mark = diamond* ,mark options=solid] table [trim cells=true,x=M,y=RT4] {data/masterRTfichera};


        \end{axis}
        \end{tikzpicture}\qquad
    \begin{tikzpicture}[scale = .6]
        \begin{axis}[
        height = 7cm,
        ymode=log,
        xmode=log,
        xlabel=number of DOFs,
        grid = major,
        ylabel=max pw. error,
        legend pos=outer north east
        ]
        \addplot[line width = 1.5pt,blue,mark = triangle*,mark size=3pt] table [trim cells=true,x=DOF2,y=B2] {data/masterBfichera};
        \addplot[line width = 1.5pt,densely dashed,brown,mark = triangle* ,mark options=solid,mark size=1pt] table [trim cells=true,x=DOF2,y=RT2] {data/masterRTfichera};


        \addplot[line width = 1.5pt,red,mark = square*] table [trim cells=true,x=DOF3,y=B3] {data/masterBfichera};
        \addplot[line width = 1.5pt,densely dashed,red,mark = square* ,mark options=solid] table [trim cells=true,x=DOF3,y=RT3] {data/masterRTfichera};
        \addplot[line width = 1.5pt,gray,mark = diamond*] table [trim cells=true,x=DOF4,y=B4] {data/masterBfichera};
        \addplot[line width = 1.5pt,densely dashed,gray,mark = diamond* ,mark options=solid] table [trim cells=true,x=DOF4,y=RT4] {data/masterRTfichera};


        \end{axis}
        \end{tikzpicture}\qquad
        \begin{tikzpicture}[scale = .6]
        \begin{axis}[
        height = 7cm,
        ymode=log,
        xmode=log,
        xlabel=GMRES iterations,
        grid = major,
        ylabel=max pw. error,
        legend pos=outer north east
        ]
        \addplot[line width = 1.5pt,blue,mark = triangle*,mark size=3pt] table [trim cells=true,x=ItB2,y=B2] {data/masterBfichera};
        \addlegendentry{$p=1$ B-spline}
        \addplot[line width = 1.5pt,densely dashed,brown,mark = triangle* ,mark options=solid,mark size=1pt] table [trim cells=true,x=ItRT2,y=RT2] {data/masterRTfichera};
        \addlegendentry{$p=1$ RT}


        \addplot[line width = 1.5pt,red,mark = square*] table [trim cells=true,x=ItB3,y=B3] {data/masterBfichera};
        \addlegendentry{$p = 2$ B-spline}
        \addplot[line width = 1.5pt,densely dashed,red,mark = square* ,mark options=solid] table [trim cells=true,x=ItRT3,y=RT3] {data/masterRTfichera};
        \addlegendentry{$p = 2$ RT}
        \addplot[line width = 1.5pt,gray,mark = diamond*] table [trim cells=true,x=ItB4,y=B4] {data/masterBfichera};
        \addlegendentry{$p=3$ B-spline}
        \addplot[line width = 1.5pt,densely dashed,gray,mark = diamond* ,mark options=solid] table [trim cells=true,x=ItRT4,y=RT4] {data/masterRTfichera};
        \addlegendentry{$p=3$ RT}


        \end{axis}
        \end{tikzpicture}
        \caption{Fichera Cube example, $\bb x_0 = [0.5,0.5,0.5]^\top$, $\kappa = 1$.}    
\label{fig::fichera}
\end{figure*}

\subsection{Example III: Toy Boat}

As a final example, we consider the Toy Boat geometry, see Fig. \ref{fig::geoms} c), where the mesh underlying the second refinement level is depicted in Fig.~\ref{fig::boat}. It consists of 28 quadratic rational Bézier patches, with extreme angles around the ``bridge'' and differences in patch size, ranging from patches of diameter $\approx 4$ to patches of diameter $\approx 0.1$, again located at the ``bridge''. This time, we compute a scattering problem with a dipole at $\bb x_0 = [7,2,0]^\top$ as a source. 
The wavenumber is $\kappa = 5$.
This induces an analytical solution on the interior which is used to verify the quality of the solution at a cluster of nine points scattered around the point $[1,0,0]^\top$. 

This third numerical example confirms what could be seen in the previous ones. Due to the complex non-smooth geometry, the orders of convergence are not as clearly visible as in the Sphere example, and cannot be easily predicted by theory. However, one still can see a positive impact from higher order approaches.
Again, since $\bb{\mathcal{RT}}\supseteq \bb\S$, one finds that the results of $\bb{\mathcal{RT}}$ achieve higher accuracies if one compares the same underlying mesh, see the left graph of Fig.~\ref{fig::boat}. However, w.r.t.~to accuracy per DOF, the spline space $\bb\S$ yields better results.
Due to the larger size of several of the patches of the geometry, and the higher wavenumber of $5$, one can expect this problem to be not as well conditioned as the other two examples. 
Due to a restart after 1500 interior iterations, this results in prohibitively high iteration numbers, especially for the $\bb{\mathcal{RT}}$ examples. This could be overcome by utilization of preconditioning or other values of accuracy and restart value of the solver. However, this is a research topic of its own, see e.g. \cite{condition}, and would exceed the scope of this article.

\begin{figure*}\centering
    \includegraphics[width=.32\textwidth]{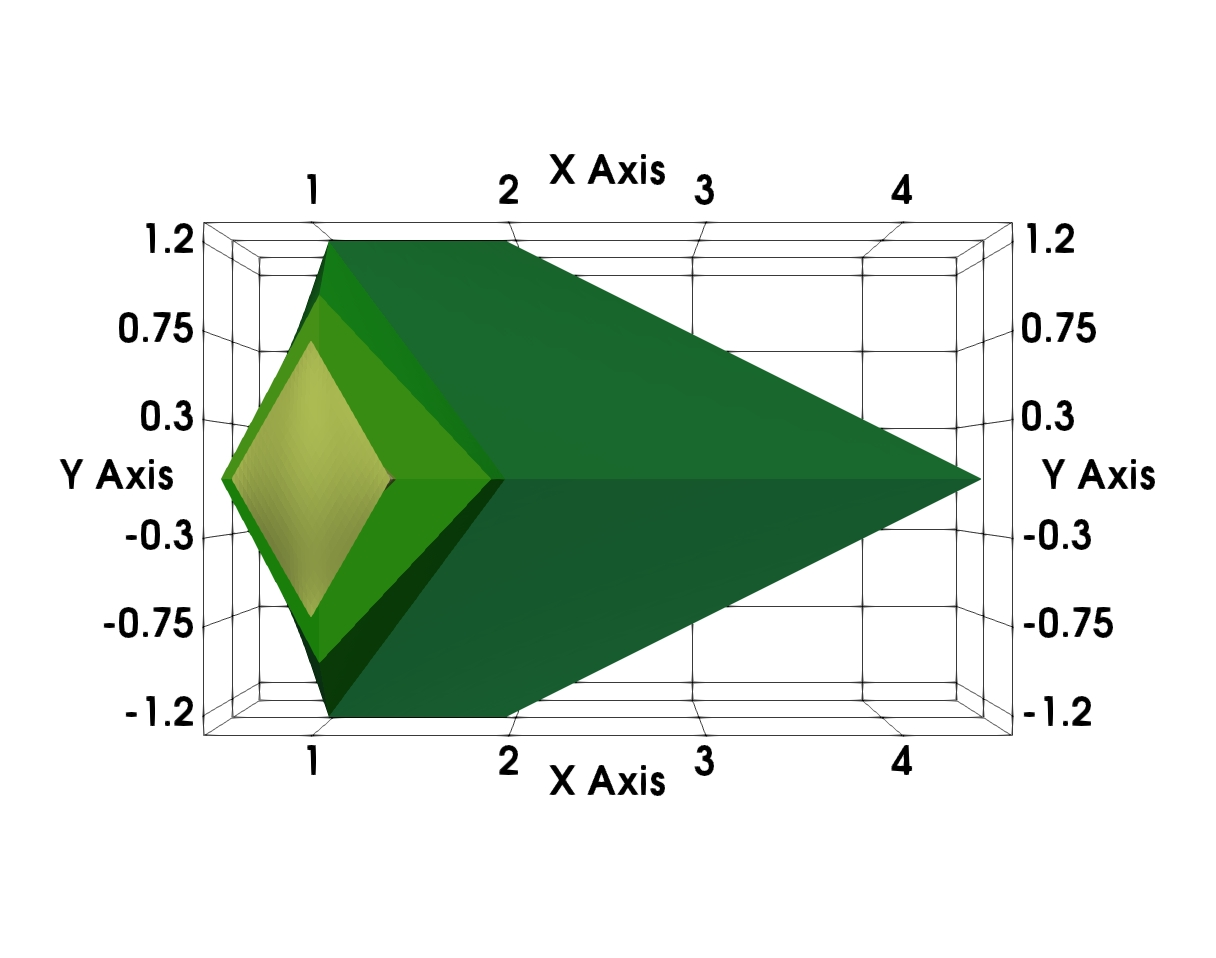}
    \includegraphics[width=.32\textwidth]{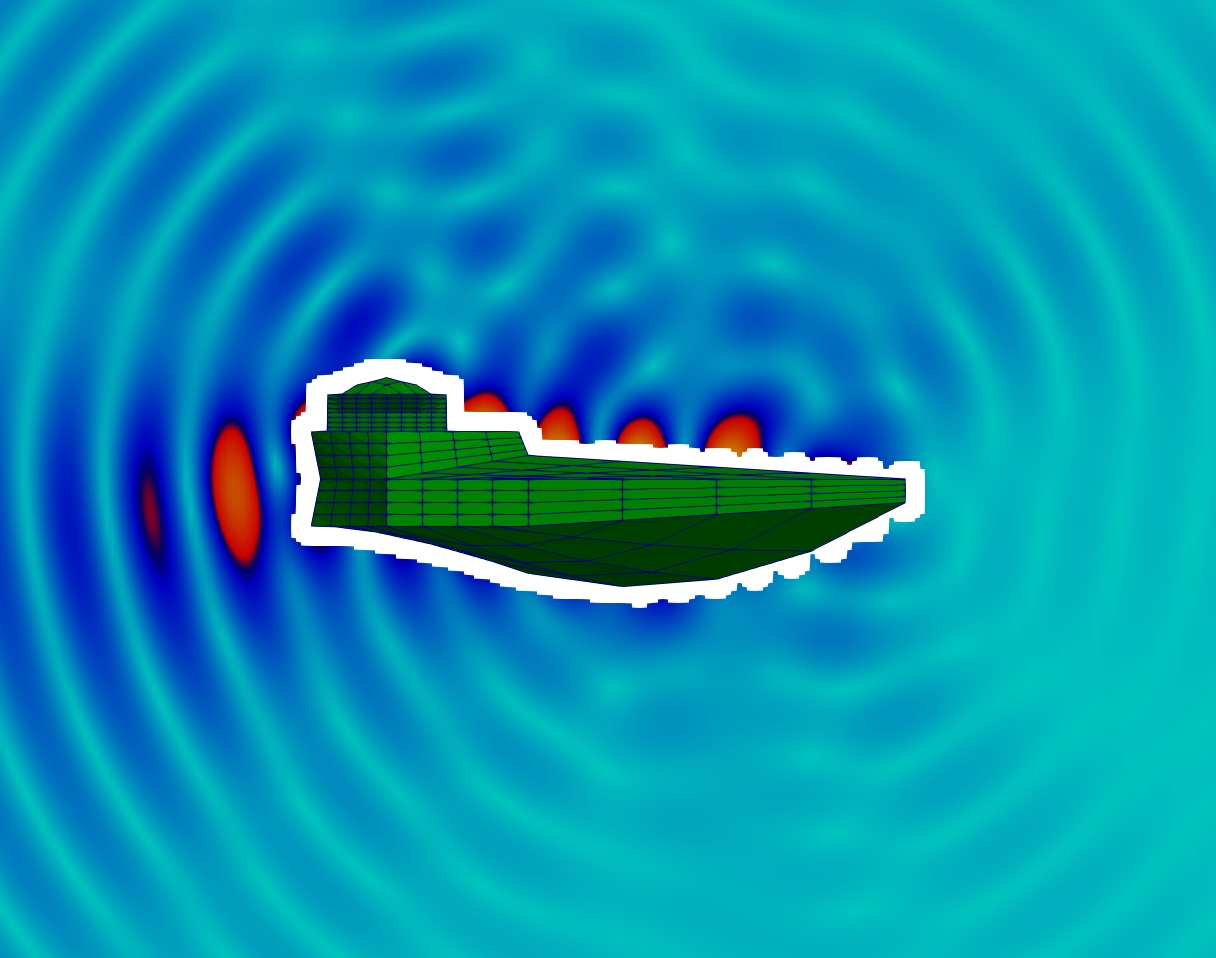}
    \includegraphics[width=.32\textwidth]{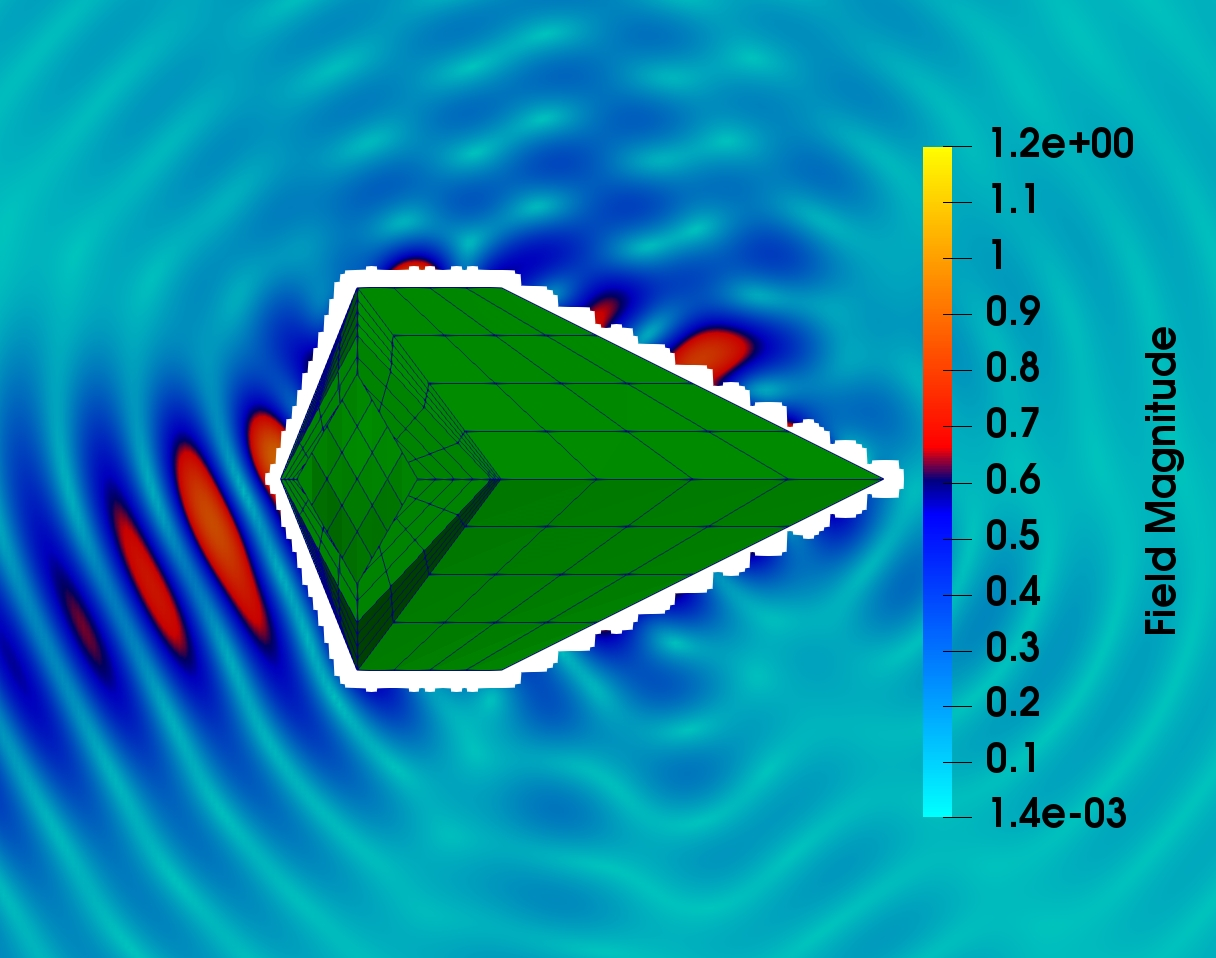}
    \caption{Toy Boat geometry and scattered field with mesh induced by refinement of level 2}\label{fig::boatpic}
\end{figure*}

\begin{figure*}[!t]
\centering
    \begin{tikzpicture}[scale = .6]
        \begin{axis}[
        height = 7cm,
        ymode=log,
        xlabel=level of refinement,
        xtick={1,2,3,4,5},
        grid = major,
        ylabel=max pw. error,
        legend pos=outer north east
        ]

        \addplot[line width = 1.5pt,red,mark = square*] table [trim cells=true,x=M,y=B2err] {data/masterShip};
        \addplot[line width = 1.5pt,densely dashed,red,mark = square* ,mark options=solid] table [trim cells=true,x=M,y=RT2err] {data/masterShip};

        \addplot[line width = 1.5pt,gray,mark = diamond* ,mark options=solid] table [trim cells=true,x=M,y=B3err] {data/masterShip};
        \addplot[line width = 1.5pt,densely dashed,gray,mark = diamond* ,mark options=solid] table [trim cells=true,x=M,y=RT3err] {data/masterShip};

        \end{axis}
        \end{tikzpicture}\qquad
 \begin{tikzpicture}[scale = .6]
        \begin{axis}[
        height = 7cm,
        ymode=log,
        xmode=log,
        xlabel=number of DOFs,
        grid = major,
        ylabel=max pw. error,
        legend pos=outer north east
        ]

        \addplot[line width = 1.5pt,red,mark = square*] table [trim cells=true,x=B2dof,y=B2err] {data/masterShip};
        \addplot[line width = 1.5pt,densely dashed,red,mark = square* ,mark options=solid] table [trim cells=true,x=RT2dof,y=RT2err] {data/masterShip};

        \addplot[line width = 1.5pt,gray,mark = diamond* ,mark options=solid] table [trim cells=true,x=B3dof,y=B3err] {data/masterShip};

        \addplot[line width = 1.5pt,densely dashed,gray,mark = diamond* ,mark options=solid] table [trim cells=true,x=RT3dof,y=RT3err] {data/masterShip};

        \end{axis}
        \end{tikzpicture}
        \qquad
 \begin{tikzpicture}[scale = .6]
        \begin{axis}[
        height = 7cm,
        ymode=log,
        xmode=log,
        xlabel=GMRES iterations,
        grid = major,
        ylabel=max pw. error,
        legend pos=outer north east
        ]

        \addplot[line width = 1.5pt,red,mark = square*] table [trim cells=true,x=ItB2,y=B2err] {data/masterShip};
        \addlegendentry{$p = 2$ B-spline}
        \addplot[line width = 1.5pt,densely dashed,red,mark = square* ,mark options=solid] table [trim cells=true,x=ItRT2,y=RT2err] {data/masterShip};
        \addlegendentry{$p = 2$ RT}

        \addplot[line width = 1.5pt,gray,mark = diamond* ,mark options=solid] table [trim cells=true,x=ItB3,y=B3err] {data/masterShip};
        \addlegendentry{$p=3$ B-spline}

        \addplot[line width = 1.5pt,densely dashed,gray,mark = diamond* ,mark options=solid] table [trim cells=true,x=ItRT3,y=RT3err] {data/masterShip};
        \addlegendentry{$p=3$ RT}

        \end{axis}
        \end{tikzpicture}
 
        \caption{Toy Boat example, $\bb x_0 = [7,2,0]^\top$, $\kappa = 5$.} \label{fig::boat}
\end{figure*}
\setlength{\tabcolsep}{5pt}

\begin{table}
  \caption{Showcase of examples of comparable accuracy}
  \label{examples}
  \begin{tabular}{cc|rrr}
  &&Example 1 & Example 2 & Example 3\\\hline
  Geometry& & Sphere &  Fichera & Toy Boat\\
  $p$&&4&3&2\\\hline
  \multirow{2}{*}{refinement level} &$\bb\S$ &3&3&4\\
   &$\bb{\mathcal{RT}}$ &2&2&4\\\hline
  \multirow{2}{*}{Number of DOFs} &$\bb\S$&2904&9600&32368\\
   &$\bb{\mathcal{RT}}$&6144&13824&114688\\\hline
  \multirow{2}{*}{Error} &$\bb\S$ &3.021e-11&1.626e-08&5.724e-05\\
  &$\bb{\mathcal{RT}}$&4.663e-11&1.809e-07&3.747e-05\\ \hline
  \multirow{2}{*}{Iterations} &$\bb\S$ &783&2505&7483\\
  &$\bb{\mathcal{RT}}$&958&4363&58169\\\hline
  \end{tabular}
\end{table}

\subsection{The Condition of the System}

It is not only of interest to compare accuracy results, but also time to solution.
Due to the superspace approach, the assembly times of the systems for B-splines and Raviart-Thomas elements virtually coincide. The effort required for the interpolation algorithm is negligible compared to the quadrature used for matrix assembly.

However, since the discrete system obtained by the projection approach is equivalent to the system obtained by a straightforward matrix assembly in the respective basis, the systems obtained by B-spline and Raviart-Thomas bases are conditioned differently.

To showcase the differences, each of the Figures \ref{fig::sphere}, \ref{fig::fichera} and \ref{fig::boat} shows the accuracy of the solution w.r.t. the number of iterations required for solving the linear system \eqref{eq::notsonaivesystem}.

The applied solver is an unpreconditioned, complex \mbox{GMRES} with a stopping criterion of a relative residual $\bb r$ with $\norm{\bb r}_{2}<10^{-10}.$
Note that, due to the increasing size of the Krylov-space, we restart the solver after every 1500 iterations.

Overall, the tendency is that, for the same polynomial degree and comparable accuracies, the B-spline systems require fewer iterations to solve. Note that, as explained above, for a given accuracy, the B-spline systems are of smaller size. While one cannot with certainty claim that the conditioning of the B-spline systems is better, a smaller system with a smaller number of iterations will yield shorter times to solution.
For comparison, three specific examples are highlighted in Table~\ref{examples}.

\section{Conclusion}
We showed that for a numerical approach via the electric field integral equation, the divergence conforming isogeometric spaces (introduced by \cite{IGADIFF}, c.f.~\cite{NP} for multipatch generalizations and estimates in trace spaces) 
admit a higher accuracy per DOF in all examples, and thus result in smaller discrete systems, both on smooth and non-smooth, non-convex geometries of different complexity. 
Both approaches were identical with the exception of the utilized basis, i.e., both approaches utilized the same geometry description and the same solver, and merely differed in the applied basis.
We stress that due to the utilization of parametric mappings even in the Raviart-Thomas-based approach, our investigation disregards any errors induced by meshing, which would favor the isogeometric approach further.
We also compared iteration numbers of both B-spline and Raviart-Thomas systems that yield solutions with comparable accuracies. Here, the tendency shows that Raviart-Thomas based systems require more effort to solve than their isogeometric counterparts. This behavior can be observed specifically for complex non-smooth geometries.
In all cases, GMRES was used without preconditioner. This lead to prohibitively large iteration numbers, showing that preconditioning is an urgent future direction of research.

\end{document}

%% file: data/deRham.tex
\begin{figure}
	\begin{tikzpicture}
		\node (center) {\hspace{.3cm}$\times$};
		\node[above of =center, node distance = 1cm] {\begin{tiny}
			\input{data/fig2aRT}	
		\end{tiny}};
		\node[below of =center, node distance = 1cm] {\begin{tiny}
			\input{data/fig2bRT}	
		\end{tiny}};
	\end{tikzpicture}
	\quad
	\begin{tikzpicture}
		\node (center) {\hspace{.3cm}$\times$};
		\node[above of =center, node distance = 1cm] {\begin{tiny}
			\input{data/fig2a}	
		\end{tiny}};
		\node[below of =center, node distance = 1cm] {\begin{tiny}
			\input{data/fig2b}	
		\end{tiny}};
	\end{tikzpicture}
	\caption{Comparison of conforming ansatz functions, for one step of interior refinement. Quadrilateral Raviart-Thomas left, spline-based right, orders $(5,4)\times(4,5)$.}
\end{figure}